\documentclass[prb,twocolumn,showpacs]{revtex4}

\usepackage{graphicx}
\usepackage[centertags]{amsmath}

\newcommand{\cm}{\ensuremath{\,\mbox{cm}^{-1}}}
\newcommand{\K}{\ensuremath{\,\mbox{K}}}
\newcommand{\celsius}{\ensuremath{\,{}^\circ}\!C}

\newcommand{\BF}{BiFeO$_{3}$}

\begin{document}

\title{Infrared and THz studies of polar phonons and improper magnetodielectric effect in multiferroic \BF\, ceramics}

\author{ S.~Kamba, D.~Nuzhnyy, M.~Savinov, J. \v{S}ebek and J.~Petzelt}
\affiliation{Institute of Physics, Academy of Sciences of the Czech Republic, Na
Slovance~2, 182 21 Prague~8, Czech Republic} \email{kamba@fzu.cz}
\author{J. Prokle\v{s}ka}
\affiliation{Charles University, Faculty of Mathematics and Physics, Department of
Condensed Matter Physics, Ke Karlovu 5, Prague 2 121 16, Czech Republic}
\author{R. Haumont}
\affiliation{Laboratoire de Physico-Chimie de l'Etat Solide - ICMMO - UMR CNRS 8182.
Universit\'{e} Paris XI, 91405 Orsay Cedex, France }
\author{J. Kreisel}
\affiliation{Laboratoire des Mat\'{e}riaux et du G\'{e}nie Physique (CNRS), Grenoble
Institute of Technology, MINATEC, 3, parvis Louis Néel, F-38016 Grenoble, France}

\date{\today}


\begin{abstract}
\BF\, ceramics were investigated by means of infrared reflectivity and time domain THz
transmission spectroscopy at temperatures 20 - 950\K\, and the magnetodielectric effect
was studied at 10 - 300\K\, with the magnetic field up to 9 T. Below 175 K, the sum of
polar phonon contributions into the permittivity corresponds to the value of measured
permittivity below 1\,MHz. At higher temperatures, a giant low-frequency permittivity was
observed, obviously due to the enhanced conductivity and possible Maxwell-Wagner
contribution. Above 200\K\, the observed magnetodielectric effect is caused essentially
through the combination of magnetoresistance and the Maxwell-Wagner effect, as recently
predicted by Catalan (Appl. Phys. Lett. \textbf{88}, 102902 (2006)). Since the
magnetodielectric effect does not occur due to a coupling of polarization and
magnetization as expected in magnetoferroelectrics, we call it improper magnetodielectric
effect. Below 175\K\, the magnetodielectric effect is by several orders of magnitude
lower due to the decreased conductivity. Several phonons exhibit gradual softening with
increasing temperature, which explains the previously observed high-frequency
permittivity increase on heating. The observed non-complete phonon softening seems to be
the consequence of the first-order nature of the ferroelectric transition.

\end{abstract}

\maketitle

\section{Introduction}

\BF\ belongs to multiferroic magnetoelectrics, because it exhibits
simultaneously ferroelectric and antiferromagnetic order. This is
known already since the beginning of the 1960's, but the interest
to this material underwent a revival after the pioneering work of
Wang at al.\cite{wang03}, who revealed the spontaneous
polarization almost by an order of magnitude higher and
substantially higher magnetization in \BF\, thin film compared to
the bulk samples. Very recently, these results were questioned,
\cite{eerenstein05} but some other experiments support the former
results.\cite{wang05} Recently many scientists paid their
attention to magnetoelectric materials not only because of the
rich and fascinating fundamental physics (see reviews
\cite{smolenskii82,fiebig05}), but also because of the promising
potential applications in multiple-state memory elements.

A ferroelectric phase transition from the cubic $Pm\bar{3}m$ to rhombohedral
$\textit{R3c}$ phase\cite{kubel90} occurs in \BF\ at T$_{C}\cong$\,1120\K, and an
antiferromagnetic ordering appears below the N\'eel temperature T$_{N}\cong$\,640\K.
\BF\, is slightly electrically conducting, which prevents to study its dielectric
properties like polarization and dielectric permittivity at room and higher temperatures.
The ferroelectric hysteresis curve is well pronounced only at low temperatures below
100\K, but the observed spontaneous polarization P$_{S}$ = 6.1.10$^{-2}$\,$\mu$C
cm$^{-2}$ is much lower than expected from the high T$_{c}$ and large lattice
distortion.\cite{teaque70} We note that smaller P$_{S}$ could be a natural consequence of
the improper nature of ferroelectricity evidenced by the doubling of the primitive unit
cell at the phase transition which was also confirmed by the first principle calculation.
\cite{wang03} Recent measurements \cite{wang03,yun04,singh06} performed on thin films
revealed P$_{S}$ by one order of magnitude higher, which was explained by a different
ferroelectric structure, namely the tetragonal structure without the cell doubling
(proper ferroelectric transition). Nevertheless, there is still an open question if it is
not an artifact due to a higher conductivity in somewhat reduced thin films, as suggested
by Eerenstein et al.\cite{eerenstein05} Temperature dependence of the bulk permittivity
was investigated above room temperature (RT) only in the high-frequency and microwave
range,\cite{roginskaya66,krainik66}, where the conductivity does not prevent the
permittivity measurements. Gradual increase in permittivity from $\sim$ 40 (at RT) to
$\sim$ 130 near T$_{C}$ was observed at 10\,GHz,\cite{krainik66}, but the permittivity
was measured only up to close above T$_{C}$.

Lattice dynamics of \BF\, was investigated only recently by means of Raman
scattering.\cite{singh05,singh06b,haumont06} It was revealed that the Raman active
phonons abruptly disappear near T$_{C}$, which supports the first-order nature of the
PT.\cite{haumont06} Phonon anomalies were discovered near T$_{N}$, but no strong phonon
softening was observed near T$_{C}$. Therefore Haumont et al.\cite{haumont06} suggested
that the ferroelectric PT in \BF\, is not soft-mode driven. Factor group analysis of the
lattice vibrations in the $R3c$ structure with two formula units per unit cell (Z=2)
yields the following optic phonons
\begin{eqnarray}
\Gamma_{R3c} = 4A_{1}(z,x^{2}+y^{2},z^{2}) + 5 A_{2}(-)+ \nonumber \\ + 9
E(x,y,x^{2}-y^{2},xy,xz,yz).
 \label{eq:rhombo}
\end{eqnarray}
It means that 4$A_{1}$ and 9$E$ modes are both Raman and infrared (IR) active, while
5$A_{2}$ modes are silent. The paraelectric $Pm\bar{3}m$ phase with Z=1 gives rise to 3
only IR active modes and one silent mode:
\begin{equation}
 \Gamma_{Pm\bar{3}m} = 3 F_{1u}(x) + 1 F_{2u}(-).
 \label{eq:cubic}
\end{equation}
This explains the vanishing of phonons from the Raman spectra
above T$_{C}$.\cite{haumont06}

IR spectra of \BF\, were not yet investigated. We found only one brief report in the
literature about IR spectra of Bi$_{1-x}$La$_{x}$FeO$_{3}$ powders measured only at room
temperature.\cite{kaczmarek75} The aim of this work is to study the temperature
dependence of polar modes, including their contribution into permittivity, and compare it
with the experimental low-frequency permittivity. We will show that the observed increase
in the intrinsic permittivity on heating is due to the gradual phonon softening without
additional dielectric dispersion between the THz and MHz range. The revealed incomplete
phonon softening towards T$_{C}$ will be assigned to the first order nature of the
ferroelectric transition. Above 175 K, the effective permittivity gains giant values due
to the enhanced conductivity and Maxwell-Wagner effect.

Influence of the magnetic field on the dielectric permittivity (magnetodielectric effect)
will be investigated up to 9 T between 10 and 300\K\, and we will show, that the
magnetodielectric effect in \BF\, is not a consequence of coupling of spontaneous
polarization and magnetization but due to combination of magnetoresistence and the
Maxwell-Wagner effect, how it was recently theoretically proposed by
Catalan.\cite{catalan06}

\section{Experimental}
\BF\ powders were prepared by conventional solid-state reaction
using high-purity (better than 99.9\%) Bi$_{2}$O$_{3}$ and
Fe$_{2}$O$_{3}$ as starting compounds. After mixing in
stoichiometric proportions, powders were calcined at 850\celsius\,
for 2h, uniaxially cold pressed, and sintered at 880\celsius\, for
2h, similarly to the synthesis proposed by Wang at
al.\cite{wang04} At the end of the procedure, we obtained almost a
pure perovskite phase of \BF. Our XRD analysis revealed only one
very tiny pattern of Bi$_{2}$O$_{3}$ giving the evidence that the
concentration of the second phase is less than 4\%. Such
concentration cannot have significant influence on the IR spectra.
The ceramics were slightly porous (less than 5\%), therefore the
reflectivity above 250\cm\, could be slightly reduced by diffuse
scattering on the surface roughness. Nevertheless, such
imperfections cannot appreciably influence the phonon frequencies
and their relative changes with temperature, which are the main
tasks of our studies.

Dielectric response of \BF\, ceramics was investigated from 10\K\,
to 700\K\, using an impedance analyzer HP 4192A (100 Hz-1 MHz).
The magnetoelectric effects were determined by measuring the
changes of permittivity and resistivity with magnetic fields up to
9 T (PPMS, Quantum design) at temperatures 10 - 300\K. The
measurements were performed at frequency 1\,kHz with
ultra-precision capacitance bridge Andeen-Hagerling 2500A. The
same ceramic disk with diameter of 7.3\,mm and thickness of
0.78\,mm was used in both studies with and without magnetic field.

Measurements at THz frequencies from 3 cm$^{-1}$ to 60 cm$^{-1}$ (0.09 - 1.8\,THz) were
performed in the transmission mode using a time-domain THz spectrometer based on an
amplified femtosecond laser system. Two ZnTe crystal plates were used to generate (by
optic rectification) and to detect (by electro-optic sampling) the THz pulses. Both the
transmitted field amplitude and phase shift were simultaneously measured; this allows us
to determine directly the complex dielectric response $\varepsilon^{\ast}(\omega)$. An
Optistat CF cryostat with thin mylar windows (Oxford Inst.) was used for measurements
down to 10\,K. For sample heating up to 900\K, we used an adapted commercial
high-temperature cell (SPECAC P/N 5850) with 1\,mm thick sapphire windows. Because of the
high THz absorption, the sample was a plane-parallel plate (diam. 7\,mm) of only
46\,$\mu$m thickness.

IR reflectivity spectra were obtained using a Fourier transform IR
spectrometer Bruker IFS 113v in the frequency range of 20 - 3000
cm$^{-1}$ (0.6 -90 THz) above RT, at lower temperature the reduced
spectral range up to 650 cm$^{-1}$ was investigated since this is
the transparency range of the polyethylene windows of our
cryostat. Pyroelectric deuterated triglycine sulfate detectors
were used for the room and higher temperature measurements, while
more sensitive He-cooled (1.5 K) Si bolometer was used for the
low-temperature measurements. Commercial high-temperature sample
cell (SPECAC P/N 5850) was used for the high-temperature
experiments up to 950\K. No windows were needed because the cell
was placed in the vacuum chamber of the spectrometer. Thermal
radiation entering the interferometer from the hot sample was
taken into account in our spectra evaluation, but it also enhanced
the noise in the spectra especially below 100\cm. Polished
disk-shaped samples with a diameter of 8\,mm and thickness of
$\sim$0.8\,mm were used.

\section{Results and evaluations}

\subsection{Infrared and THz studies}

\begin{figure}
\begin{center}
    \includegraphics[width=83mm]{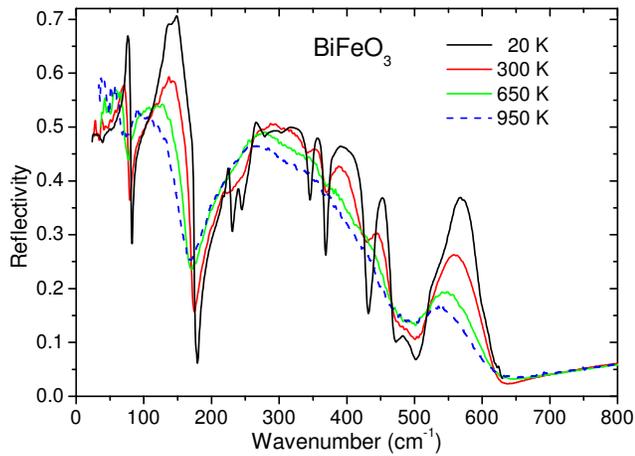}
  \end{center}
\caption{(Color online) Temperature dependence of the IR
reflectivity spectra of \BF\, ceramics. We note that the
reflectivity value above 200\cm\, may be slightly reduced due to a
small porosity of the ceramics and subsequent diffuse scattering
of the IR beam. This can apparently enhance the phonon damping in
the fit of our spectra, but the phonon frequencies are not
substatially influenced.} \label{Fig1}
\end{figure}

Experimental IR reflectivity spectra of \BF\ plotted at selected
temperatures between 20 and 950\K\, are shown in Fig.~\ref{Fig1}.
13 reflection bands are well resolved at 20\K, which exactly
agrees with the predicted number of IR active modes in the
rhombohedral phase (see Eq~\ref{eq:rhombo}). Most of the phonons
gradually weaken on heating because the strengths of the newly
activated modes in the rhombohedral phase are proportional to the
square of the order parameter \cite{petzelt76}. Simultaneously
dampings of all modes increase on heating. Both these effects
cause that apparently only four IR reflection bands are resolved
at 950\K\, (see Fig.~\ref{Fig1}) although actually still 13 modes
are needed for the reflectivity fit up to the highest temperature
(see the fitting method below). One can expect that the strength
of most modes will further gradually decrease on heating above
950\K\, and stepwise vanish at T$_{c}$ $\cong$ 1120\K\, due to the
first order phase transition into the cubic phase, where only 3
polar modes are permitted by symmetry.

\begin{figure}
\begin{center}
    \includegraphics[width=85mm]{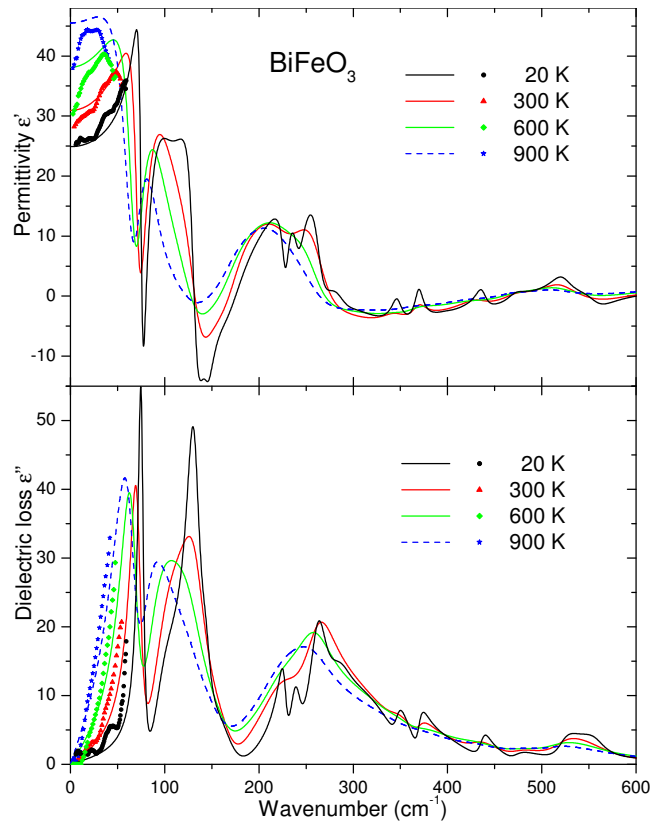}
  \end{center}
\caption{(Color online) Complex dielectric spectra of \BF\, at
selected temperatures. The dots are experimental THz data, lines
are results of the reflectivity fits. Note the shifts of
$\varepsilon$'' peaks to lower frequencies due to the phonon
softening on heating.} \label{Fig2}
\end{figure}
IR and THz spectra were fitted simultaneously using a generalized-oscillator model with
the factorized form of the complex permittivity:\cite{gervais83}
\begin{equation}\label{eps}
\varepsilon^{*}(\omega)=\varepsilon_{\infty}\prod_{j}\frac{\omega^{2}_{LOj}-\omega^{2}+i\omega\gamma_{LOj}}{\omega^{2}_{TOj}-\omega^{2}+i\omega\gamma_{TOj}}
\end{equation}
where $\omega_{TOj}$ and $\omega_{LOj}$ denotes the transverse and
longitudinal frequency of the j-th polar phonon, respectively, and
$\gamma$$_{TOj}$ and $\gamma$$_{LOj}$ denotes their corresponding
damping constants. $\varepsilon$$^{*}$($\omega$) is related to the
reflectivity R($\omega$) by
\begin{equation}\label{refl}
R(\omega)=\left|\frac{\sqrt{\varepsilon^{*}(\omega)}-1}{\sqrt{\varepsilon^{*}(\omega)}+1}\right|^2
.
\end{equation}
The high-frequency permittivity $\varepsilon_{\infty}$ resulting
from the electronic absorption processes was obtained from the
room-temperature frequency-independent reflectivity tails above
the phonon frequencies and was assumed temperature independent.

\begin{figure}
    \begin{center}
      \includegraphics[width=80mm]{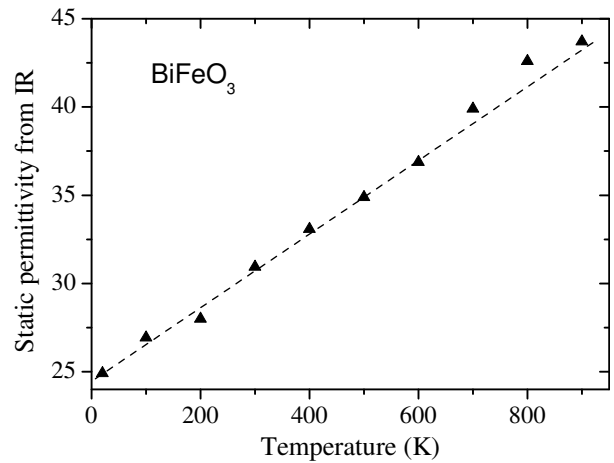}
    \end{center}
\caption{Temperature dependence of the static permittivity $\varepsilon_{0}$ obtained
from the fit to IR reflectivity.\label{Fig3}}
\end{figure}

Real and imaginary parts of $\varepsilon^*$($\omega$) obtained
from the fits to IR reflectivity are shown together with the
experimental THz spectra in Fig.~\ref{Fig2}. The high-temperature
THz data do not exactly correspond to the fit of reflectivity
mainly due to possible inaccuracy of the THz experiment as a
consequence of possibly different temperature of the sapphire
windows in the furnace during the separate reference and sample
measurements. Nevertheless, one can see the frequency shift
(softening) in the maxima of dielectric-loss spectra
($\varepsilon$'') with increasing temperature. Since the
$\varepsilon$''($\omega$) maxima correspond to $\omega_{TOj}$
frequencies (for not too-heavily damped modes), one can see that
most of the polar phonons soften on heating. The phonon softening
causes an gradual increase in the static permittivity
$\varepsilon_{0}=\sum {\Delta\varepsilon_{j}} +
\varepsilon_{\infty}$ with increasing temperature (see
Fig.~\ref{Fig3}) because the sum $f$ of all the oscillator
strengths $f_{j}$ is expected to be practically temperature
independent:
\begin{equation}
 f(T)=\sum_{j=1}^n{f_j}=\sum_{j=1}^n{\Delta\varepsilon_{j}.\omega_{TOj}^2} = const.
 \label{eq:sum_f}
\end{equation}
$\Delta\varepsilon_{j}$ denotes the contribution of the j-th mode
to static permittivity and can be obtained from the
formula\cite{gervais83}
\begin{equation}
 \Delta\varepsilon_{j} = \varepsilon_{\infty}\omega^{-2}_{TOj}\frac{\prod_{k}\omega^{2}_{LOk}-\omega^{2}_{TOj}}{\prod_{k\neq
 j}\omega^{2}_{TOk}-\omega^{2}_{TOj}}.
 \label{eq:sila}
\end{equation}
Eq.~\ref{eq:sum_f} is expected to be fulfilled on very general
basis of summation rules\cite{smith85} and in the case of
uncoupled phonons even each oscillator strength $f_{j}$ remains
temperature independent (i.e.
$\Delta\varepsilon_{j}.\omega_{TOj}^2 = const.$). It means that if
e.g. the j-th phonon frequency reduces twice its frequency
$\omega_{TOj}$, its dielectric strength $\Delta\varepsilon_{j}$
increases four times. Partial softening of two lowest frequency
modes from 76 and 99\cm\, at 20 K to 67 and 82\cm\, at 900 K,
respectively, explains the observed rise in permittivity on
heating (Fig.~\ref{Fig3}) similarly as the Cochran-type softening
explains the Curie-Weiss anomaly near the ferroelectric PT in
proper displacive ferroelectrics. Fig.~\ref{Fig3} shows that the
static permittivity $\varepsilon_{0}$ from phonon contributions
continuously increases almost linearly with temperature.
Experimental value of $\varepsilon_{0}$ obtained above 600\K\,
slightly deviates from the linear fit (see Fig.~\ref{Fig3}), but
it is difficult to claim that it is due to spin-phonon interaction
which disappears above the N\'{e}el temperature (T$_{N}$ $\approx$
640\K), because the deviation from the linear fit lies within the
limits of our experimental accuracy. Nevertheless, there are some
other experimental evidences (e.g. measurement of temperature
dependence of the rhombohedral angle around
T$_N$)\cite{palewicz06} suggesting a strong spin-lattice coupling.

Let us compare the static permittivity $\varepsilon_{0}$ in Fig.~\ref{Fig3} with the
published 300 MHz\cite{roginskaya66} and 10 GHz\cite{krainik66} dielectric data. Our
values of $\varepsilon$'(T) in Fig.~\ref{Fig3} correspond to 300 MHz
data.\cite{roginskaya66} Krainik at al.\cite{krainik66} observed slightly higher
$\varepsilon$'(T), but the 10 GHz high-temperature experiment is rather complicated and
of lower accuracy, therefore we believe more to the 300 MHz data, which correspond well
to ours. Small dielectric anomaly seen at T$_{C}$ in Ref. \cite{krainik66} is compatible
with the improper ferroelectric nature of the PT due to the doubling of unit cell
\cite{kubel90} below T$_{C}$. The non complete phonon softening of two lowest frequency
modes (obtained from our IR spectra taken below 950\K\, by extrapolation of phonon
frequencies to T$_{C}$) can be explained by the probably strongly first-order nature of
the ferroelectric transition in \BF.

\begin{table}
\caption{\label{tab:table1} Polar mode parameters of the \BF\
ceramics obtained from the fit to the 300\,K reflectivity (see
Eqs.~\ref{eps} and ~\ref{refl}) and phonon frequencies observed in
Raman spectra of single crystal and thin film. All parameters are
in \cm, $\varepsilon_{\infty}$ = 4.0. Assignment of the phonon
symmetry is given in the last column.}

\begin{ruledtabular}
\begin{tabular}{cccclcc}
 \multicolumn{4}{c}{infrared spectra}&\multicolumn{2}{c}{Raman spectra} & \\
  $\omega_{TO}$ & $\gamma_{TO}$ & $\omega_{LO}$ & $\gamma_{LO}$ & crystal\cite{haumont06} &
  film\cite{singh06b} & symmetry\\
\hline

\hline
  71.9 & 14.3 & 77.6  & 7.0 & 56.1, 84.1 &  & E \\
  99.4 & 57.1 & 122.3 & 91.6 & 95.3 &       & E \\
 134.6 & 34.1 & 152.0 & 53.1 & 127.1, 141.4 & 136 & A$_1$  \\
 170.3 & 44.2 & 171.4 & 11.8 & 162.9 & 168  & A$_1$ \\
 228.0 & 52.9 & 236.3 & 45.6 & 204.6 & 212  & A$_1$\\
 262.8 & 43.4 & 272.0 & 81.7 & 261.3 & 275  & E \\
 310.5 & 164.4& 342.5 & 30.6 & 316.6 & 335  & E \\
 345.1 & 33.1 & 367.4 & 14.8 &       & 363  & E \\
 369.2 & 13.2 & 426.4 & 32.5 & 383.6 &      & E \\
 433.1 & 34.3 & 466.7 & 25.2 &       & 425  & A$_1$ \\
 472.5 & 34.6 & 499.6 & 50.5 &       & 456  & E \\
 521.1 & 45.7 & 532.8 & 116.1 & 530.9 & 549 & E \\
 554.9 & 73.3 & 603.5 & 57.7 &       & 597  & E \\
            \end{tabular}
 \end{ruledtabular}
\end{table}

Let us compare the phonon frequencies observed at RT IR spectra of
ceramics with Raman modes observed in single
crystal\cite{haumont06} and thin film\cite{singh06b} (see
Table~\ref{tab:table1}). Haumont et al.\cite{haumont06} observed
two modes in Raman spectra below 85\cm, while in IR spectra we
clearly see only one mode near 72\cm. We think that the
lower-frequency Raman mode might be artifact of the filter
removing the elastic peak from the Raman spectra. All TO modes
predicted from the factor group analysis are seen in our IR
spectra, but some of the modes are missing in Raman spectra,
presumably due to their low Raman activity. The Raman mode
frequencies in thin films are slightly shifted against the modes
seen in single crystal, possibly due to a strain effect in the
film. Polarized Raman spectra of the thin film allowed to assign
the symmetry of some modes,\cite{singh06b} and our assignment in
table I is extended to all the observed modes. The
lowest-frequency E and A$_1$ modes probably represent the soft
mode doublet stemming from the triply degenerate soft mode from
the Brillouin zone boundary (IR and Raman inactive) in the cubic
phase.

Raman modes do not exactly correspond to the IR TO mode
frequencies. It is known that the grain boundaries in ceramics may
cause stiffening of the soft mode in comparison with single
crystal due to a small-permittivity grain-boundary layer (so
called dead layer), but such effect is remarkable only in the case
of high-permittivity materials (i.e. with a strong soft mode at
low frequencies). Typical example of such ceramics, where the
stiffening of the soft mode was observed, is
SrTiO$_{3}$.\cite{petzelt06} Using effective medium approximation,
Rychetsky and Petzelt have shown\cite{rychetsky04} that all polar
phonon frequencies should exhibit an increase with the grain
boundary concentration, and the shift of the TO phonon frequency
squared should be proportional to the dielectric strength of the
mode. Since the usual polar modes have much lower dielectric
strengths than the soft mode, their shift should be much lower
than that of the soft mode. The phonons contribute less than 50 to
the relative permittivity of \BF\, while in SrTiO$_{3}$ they
contribute more than 20000 at low temperatures. Therefore the
effect of the mode stiffening due to the grain boundaries in \BF\,
should be small and insignificant for explanation of differences
between the TO phonon frequencies in ceramics and single crystals.

The problem of the mode frequency determination in the case of dielectrically anisotropic
grains (in ceramics or polycrystalline films) or domains (in polydomain crystals) is more
relevant in our case of \BF\,. The proper way how to evaluate the IR reflectivity spectra
in such a case is also through using the effective medium approximation, if the grain
(domain) size is much smaller than the probing wavelength. \cite{pecharroman94,hlinka06}
This approach may effectively shift the TO mode frequencies and may cause also some
spurious peaks (so called geometrical resonances) in the reflectivity. In our case of
standard fits using the Eq.~\ref{eps}, which neglect this problem, the determined TO
frequencies may slightly differ from the actual E and A$_{1}$ mode frequencies. Moreover
it becomes clear that the corresponding LO frequencies have no real physical meaning at
all (neglecting damping, they correspond to zeros of the effective dielectric function,
but not to zeros of the dielectric functions along the principal axes of the dielectric
ellipsoid), as well as the mode strengths and probably also the dampings. On the other
hand, also in Raman spectra the peaks do not correspond necessarily accurately to TO and
LO frequencies, but may lay in between due to the angular dispersion of the polar modes.
In this way, some differences between the evaluated IR and Raman modes in polycrystalline
samples with dielectrically anisotropic grains are quite naturally to be expected.

\subsection{Dielectric and magnetodielectric studies}

\begin{figure}
    \begin{center}
      \includegraphics[width=87mm]{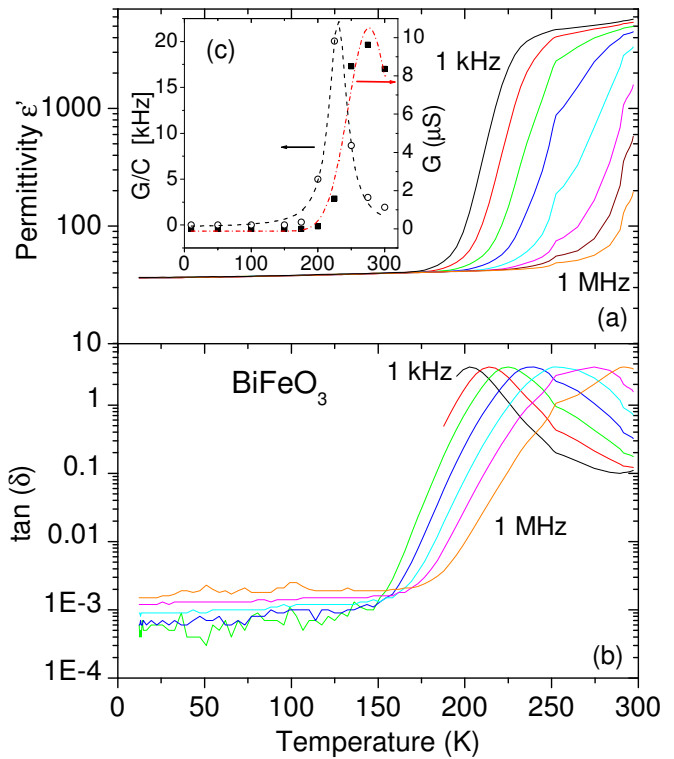}
    \end{center}
\caption{(Color online) Temperature dependence of the low-frequency permittivity (a) and
dielectric loss (b) in \BF\, ceramics. Low-frequency $\tan\delta$ is not shown at low
temperatures due to the high noise. Inset (c) shows the temperature dependence of
conductance $G$ (full symbols) and ratio of conductance and capacitance $G/C$ (empty
symbols). \label{Fig4}}
\end{figure}

We measured also the complex permittivity in the range below
1\,MHz, however the results show the intrinsic permittivity only
below 170\K\, (see Fig.~\ref{Fig4}). At higher temperatures both
$\varepsilon$' and $\tan\delta$ remarkably rise, presumably due to
a Maxwell-Wagner-type contribution to the
permittivity\cite{cohen03,liu05} as a consequence of the increased
conductivity and its inhomogeneity in the sample (see
Fig.~\ref{Fig4}c). Intrinsic $\varepsilon$' with the value below
40 is dispersion-less at low temperatures and $\varepsilon$'
slightly increases on heating due to the above mentioned phonon
softening. Within the accuracy of measurement, the value of
$\varepsilon$' corresponds to Fig.~\ref{Fig3} as well as to
previously published high-frequency
data.\cite{roginskaya66,krainik66} It means that no dielectric
relaxation is expected between the kHz and THz range below 175\K.
Some authors observed twice higher $\varepsilon$' in \BF\,
ceramics\cite{jun05} and thin films\cite{yun04,uchida06} at RT,
but this might be influenced by a Maxwell-Wagner contribution or
by the substrate induced strain in the thin film.

\begin{figure}
      \centerline{\includegraphics[width=80mm]{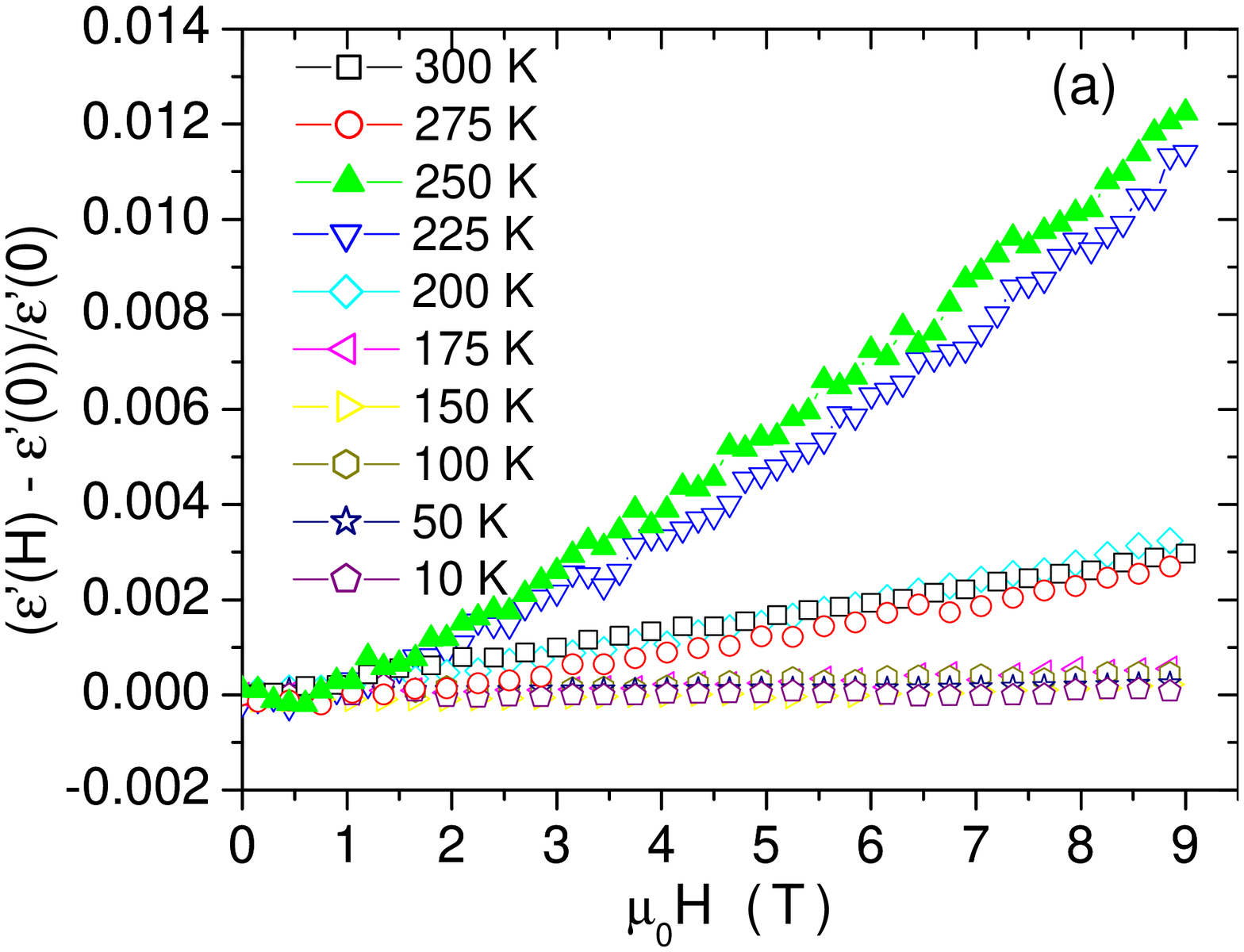}}
      \vskip 15pt
      \centerline{\includegraphics[width=80mm]{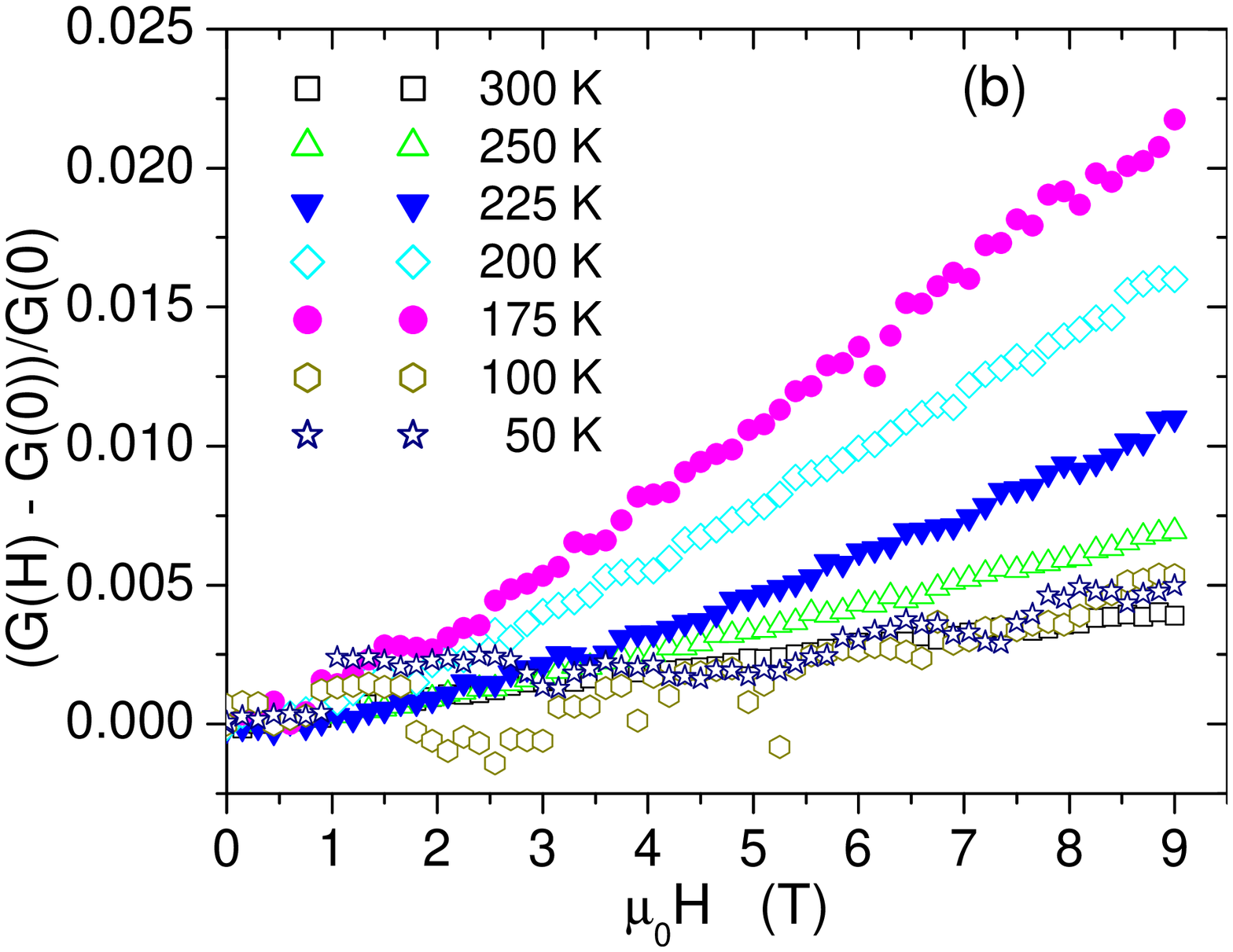}}
\caption{(Color online) Magnetic field dependence of a) relative
permittivity and b) relative conductance changes measured at
1\,kHz at various temperatures. Higher noise in conductance
changes at low temperatures is caused by the low conductivity
below 150\K. \label{Fig5}}
\end{figure}

\begin{figure}
\begin{center}
      \includegraphics[width=87mm]{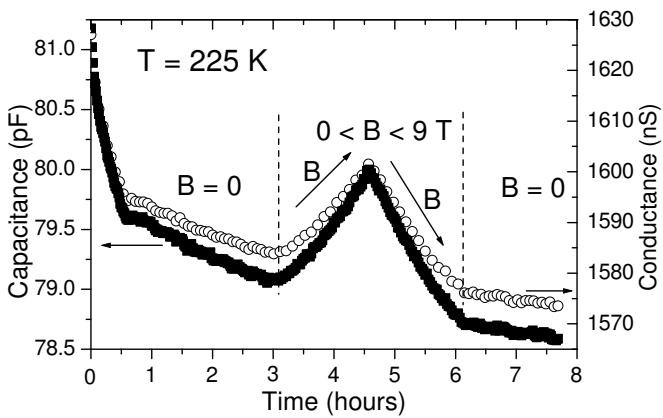}
    \end{center}
\caption{Time dependence of the capacitance (solid points) and
conductance (open points) with magnetic field continuously
increasing from 0 to 9 T and then decreasing to 0 T. \label{Fig6}}
\end{figure}

 Magnetoferroelectric materials should exhibit changes of the spontaneous polarization
and permittivity with magnetic field.\cite{smolenskii82} Nevertheless, it was found that
the linear magnetoferroelectric effect should not take place in \BF\, due to its
cycloidal antiferromagnetic structure of the G-type. The linear magnetoferroelectric
effect was observed only above 20 T, due to unwinding of the cycloidal magnetic ordering
in high magnetic fields.\cite{popov93} However, Fig.~\ref{Fig5}a shows some non-zero
variation of the permittivity with magnetic field up to 9\,T. No giant changes (like in
TbMnO$_{3}$\cite{kimura03}) were observed, but the changes at 250 and 225\K\, are one
order of magnitude larger than in recently studied Nb-doped \BF\, ceramics,\cite{jun05}
which exhibits six orders of magnitude lower conductivity than our \BF\, ceramics.
However, above 250 and below 200\K\, the magnetodielectric effect dramatically decreases
and below 175\K, i.e. in the temperature range where the Maxwell-Wagner mechanism does
not contribute to the permittivity, the magnetodielectric effect is very low
($\frac{\varepsilon(H)-\varepsilon(0)}{\varepsilon(0)}\approx$ 10$^{-5}$).
Simultaneously, Fig.~\ref{Fig5}b shows that the conductance of \BF\, is dependent on the
magnetic field. The changes are not very large, but since we measure the effective
conductance, from our measurement at a single frequency we cannot distinguish the
conductance of grain bulk and grain boundaries (as well as their changes with magnetic
field), which probably differ. In this case the impedance spectroscopy\cite{west97} in a
wide frequency and temperature range should be used, which will be the subject of our
next study.

It becomes clear that the strong magnetodielectric effect seen in
Fig.~\ref{Fig5}a above 200K\, is a consequence of a combination of
the magnetoresistance and Maxwell-Wagner effect, as recently
proposed by Catalan.\cite{catalan06} In his model it is assumed
that the grain boundaries and/or interfacial layers between the
sample and electrodes may show different resistivity and
magnetoresistance than the grain bulk. This model predicts the
maximal magnetodielectric effect around the frequency $f =
1/RC=G/C$ (R - resistivity, C - capacity, G -
conductance).\cite{catalan06} Fig.~\ref{Fig4}c shows the $G/C$
maximum of 22 kHz at 228\K\, in good agreement with the maximal
permittivity change with magnetic field observed at 1 kHz between
225 and 250\K\, (Fig.~\ref{Fig5}a).

Our explanation of the observed magnetodielectric effect in \BF\, shows that the change
of permittivity with magnetic field occurs in \BF\, not due to coupling of spontaneous
polarization and magnetization, as expected in magnetoferroelectrics, but due to a
combination of the magnetoresistance and Maxwell-Wagner effect. Therefore we suggest to
call this effect improper magnetodielectric effect. Similar effect can be seen also in
other non-ferroelectrics magnetoresistive materials.

We have to stress that our data on magnetodielectric effect were
obtained always after 3 hours of temperature stabilization,
because a slow temporal relaxation of capacitance and conductance
was unexpectedly observed at all investigated temperatures. Such
an aging deteriorated the magnetodielectric measurements if we
started only several minutes after setting the measurement
temperature. The aging was most remarkable above 200\K, i.e. in
the temperature range where the Maxwell-Wagner effect becomes
important. Example at 225\K\, is seen in Fig.~\ref{Fig6}, which
shows that even two relaxation processes contribute to the
temporal change of capacitance and conductance. The first fast
process causes the steep decrease in capacity during the first
half an hour, while the second process is remarkable even after 7
hours and the relaxation continues also after applying the
magnetic field. One could e.g. speculate that the two relaxations
are caused by two types of interfaces (grain boundaries and
dielectric-electrode interfaces). However, the question arises if
the temperature drifts cannot influence such a behavior. As the
conductance steeply changes by three orders of magnitudes between
275 and 175\K\, (from 9.6 $\mu$S to 6.1 nS - see
Fig.~\ref{Fig4}c), some small change of C or G could be expected
if the sample temperature is drifting in the mK scale. However,
the sample was placed in the He gas, which provides a good thermal
contact and therefore no temperature drifts are expected in the
long-time scale of our measurements. But we cannot exclude that
the faster relaxation (first half an hour) is due to the
temperature stabilization. Slow-time relaxation should be assigned
to some slow diffusion of defect charges, but detailed
understanding is missing. It is worth to note that some long-time
relaxations of capacitance and resistivity were observed also in
other systems.\cite{nalbach04,skrbek96} A study of these phenomena
is in progress.

\section{Conclusion}
THz and IR spectra obtained between 20 and 950\K\, revealed a
remarkable lattice softening, which explains the experimentally
observed increase in permittivity on heating. The number of
observed polar phonons corresponds to that predicted by the factor
group analysis. Strengths of the most modes gradually decrease on
heating, because only 3 polar modes are permitted in the cubic
phase above 1120\K. IR phonon frequencies were compared with the
Raman spectra and the phonon symmetries were assigned. Possible
differences between the IR and Raman frequencies are discussed.
Non-complete phonon softening towards T$_{C}$ was observed and
explained by the first-order nature of the ferroelectric
transition. The observed magnetodielectric effect and giant
low-frequency permittivity at temperatures above 200\K\, was
explained by combination of the magnetoresistance and
Maxwell-Wagner effect. Unexpected slow temporal relaxation of
capacitance and conductance was observed.

Finally it should be stressed that the magnetodielectric effect in \BF\, ceramics is not
caused by a coupling of polarization and magnetization as expected for
magnetoferroelectric multiferroics. Since it is caused by combination of
magnetoresistance and Maxwell-Wagner polarization effect, we call it improper
magnetodielectric effect. Similar magnetodielectric effect can be expected also in
nonferroelectric slightly conducting materials with large magnetoresistance.

\begin{acknowledgments}
The work was supported by the Grant Agency of the Czech Republic (Projects No.
202/06/0403, 106/06/0368 and AVOZ10100520). The authors would like to thank P. Ku\v{z}el
and S. Denisov for the technical help with the experiments and to J. Hejtm\'anek for
fruitful discussions.
\end{acknowledgments}

\end{document}